\documentclass{book}
\usepackage[latin9]{inputenc}
\usepackage{geometry}
\usepackage{verbatim}
\usepackage{wrapfig}
\usepackage{amsthm}
\usepackage{amsmath}
\usepackage{amssymb}
\usepackage{graphicx}
\usepackage{bibentry}
\usepackage{color}
\usepackage{graphicx}
\usepackage{authblk}

\makeatletter

\newcommand{\GG}{{\mathcal{G}}}

\newcommand{\NN}{{\mathcal{N}}}

\newcommand{\diag}[1]{\operatorname{diag}\left( #1\right)}

\renewcommand{\epsilon}{\varepsilon}

\renewcommand{\hat}{\widehat}

\newcommand{\Nin}{\NN^\text{in}}

\newcommand{\oprocendsymbol}{\hbox{$\bullet$}}
\newcommand{\oprocend}{\relax\ifmmode\else\unskip\hfill\fi\oprocendsymbol}

\newcommand{\longthmtitle}[1]{\mbox{}\textup{\textbf{(#1)}}}

\theoremstyle{plain}
\newtheorem{thm}{\protect\theoremname}
\theoremstyle{definition}
\newtheorem{problem}[thm]{\protect\problemname}
  \newtheorem*{problem*}{\protect\problemname}

\usepackage{amsfonts}
\usepackage{amsmath}
\usepackage{amssymb}
\usepackage{epsfig}
\usepackage{graphics}
\usepackage{graphicx}
\usepackage{needspace}
\usepackage{color, colortbl}
\definecolor{Gray}{gray}{0.9}
\epsfclipon
\newtheorem{theorem}{Theorem}[section]

\newtheorem{corollary}[theorem]{Corollary}
\newtheorem{lemma}[theorem]{Lemma}

\newtheorem{definition}{Definition}[section]

\newtheorem{proposition}{Proposition}
\newtheorem{remark}{Remark}[section]

\makeatother

\providecommand{\problemname}{Problem}

\providecommand{\theoremname}{Theorem}

\makeatother

\providecommand{\problemname}{Problem}
\providecommand{\theoremname}{Theorem}

\date{}

\begin{document}

\title{Bio-Inspired Framework for Allocation of Protection Resources in
Cyber-Physical Networks}

\author[1]{Victor M. Preciado}
\author[1]{Michael Zargham}
\author[2]{Chinwendu Enyioha}
\author[1]{Cameron Nowzari}
\author[1]{Shuo Han}
\author[1]{Masaki Ogura}
\author[3]{Ali Jadbabaie}
\author[1]{George Pappas}

\affil[1]{Department of Electrical and Systems Engineering. University of Pennsylvania.}

\affil[2]{Department of Electrical Engineering. Harvard University.}

\affil[3]{Institute for Data, Systems, and Society. Massachusetts Institute of Technology.}

\maketitle

\chapter{Introduction}
Understanding spreading processes in complex networks and designing
control strategies to contain them are relevant problems in many different
settings, such as epidemiology and public health \cite{Bai75}, computer
viruses \cite{GGT03}, information propagation in social networks \cite{LAH07}, or security of cyberphysical networks \cite{roy2012security}. In this chapter, we describe a bio-inspired framework for optimal
allocation of resources to prevent spreading processes in complex cyber-physical networks. Our motivation is inspired by recent advancement on the problem of containing epidemics in human contact networks. The most popular dynamic epidemic model is the Susceptible-Infected-Susceptible (SIS)
model~\cite{RMA-RMM-BA:92,MJK-PR:07}. In this model, a given population is divided into two compartments. The first compartment, called `Susceptible' ($S$), contains individuals who are healthy, but susceptible to becoming infected. The second compartment is called `Infected' ($I$) and contains individuals who are infected and able to recover from the disease. Individuals can transition from $S$ to $I$ as they become infected, and from $I$ to $S$ as they recover. In addition to the SIS model, there are many other models able to model more realistic spreading processes. This is often done by adding extra compartments representing a variety of disease stages. There are many works that analyze different variations of the SIS model, such as extensions to higher number of disease states~\cite{PD-JW:02,SF-EG-VAAJ:10,IZK-JC-MR-PLS:10,NP-DB-BG-AV:11,PP-BC-MA-AP-SM:09,DB-OD-WFG-AP-RV:12,MMH:14}, or explicit modeling of birth and mortality rates~\cite{HWH:00,JL-YY-YZ:11}. Stability results are obtained in~\cite{AK-GCW:02,AL-LO-DK:11,JL-YY-YZ:11} using Lyapunov analysis, or in~\cite{HWH-HWS-PVDD:81} using Volterra integral models.

In the literature, we find several approaches to model spreading mechanisms
in arbitrary contact networks. The analysis of this question in arbitrary (undirected) contact networks was first studied by Wang et al. \cite{WCWF03} for a Susceptible-Infected-Susceptible
(SIS) discrete-time model. In \cite{GMT05}, Ganesh et al. studied
the epidemic threshold in a continuous-time SIS spreading processes.
In both continuous- and discrete-time models, there is a close connection
between the speed of the spreading and the spectral radius of the
network (i.e., the largest eigenvalue of its adjacency matrix) \cite{MOK09}. Designing strategies to contain spreading processes in networks is
a central problem in public health and network security. In this context,
the following question is of particular interest: given a contact
network (possibly weighted and/or directed) and resources that provide
partial protection (e.g., vaccines and/or antidotes), how should one
distribute these resources throughout the network in a cost-optimal
manner to contain the spread? This question has been addressed
in several papers. Cohen et al. \cite{cohen2003efficient} proposed
a heuristic vaccination strategy called \emph{acquaintance immunization
policy} and proved it to be much more efficient than random vaccine
allocation. In \cite{BCGS10}, Borgs et al. studied theoretical
limits in the control of spreads in undirected network with a non-homogeneous
distribution of antidotes. Chung et al. \cite{chung2009distributing}
studied a heuristic immunization strategy based on the PageRank vector
of the contact graph. In the control systems literature, Wan et al. proposed in \cite{YW-SR-AS:07,WRS08} a method to design control strategies by allocating heterogeneous resources in undirected networks. In \cite{VMP-AJ:09}, the authors present an spectral analysis of proximity random graphs with applications to virus spread. In \cite{GOM11}, the authors study the problem of minimizing the level of infection in an undirected network using corrective resources within a given budget. In \cite{PDS13} a linear-fractional optimization
program was proposed to compute the optimal investment on disease awareness over the nodes of a social network to contain a spreading process. In particular, we will cover in detail the work in \cite{PZEJP13,cones14,PZ13,VMP-MZ-DS:14},
where the authors developed a convex formulation to find the optimal allocation
of protective resources in a network. An analysis of greedy control strategies and worst-case conditions was presented in  \cite{MZ-VMP:14}. Recent extensions include the analysis of more general epidemic models \cite{NPP15}, competing diseases \cite{CP14,WNPP15,WNPP16a}, time-switching networks \cite{OP15TV,OP16b,OP16c}, and non-Poissonian spreading and recovery rates \cite{OP15Phase,OP16a} have been recently developed. A novel data-driven optimization framework has also been recently proposed by Han et al. in \cite{HPNP15}. A distributed framework for optimal allocation of resources has also been proposed in \cite{ER-SM:14}. A novel analysis of epidemic models in arbitrary graphs using tools from positive systems can be found in \cite{KBG14}.

In this Chapter, we describe an optimization-based framework to find the optimal allocation of protection resources in \emph{weighted and directed} networks of \emph{nonidentical} agents in polynomial time. In our study, we consider two types of containment resources:
\begin{itemize}
\item \emph{Preventive} resources able to protect (or `immunize') nodes
against the spreading (such as vaccines in a viral infection process). This type of resources are allocated in nodes and/or edges of the network before the spread has reached them, so that this element is protected from the spread. The effect of this resource is to reduce the rate in which the spread can reach this element.

\item \emph{Corrective} resources able to neutralize the
spreading after it has reached a node (such as antidotes in a viral
infection). Notice that, in contrast with preventive resources, corrective resources are used after the spread has reached a node in the network. The effect of this type of resource is to increase the rate of recovery of an elements after the spread has reached it.

\end{itemize}

In the framework herein presented, we assume there are cost associated with these resources
and study the problem of finding the cost-optimal distribution of
resources throughout the network to contain the spreading.
The aforementioned protection resources have an associated cost that
depends on the level of protection achieved by the resource. For example,
the larger the investment on vaccines and antidotes, the higher the
level of protection achieved by the population in which the resources
have been distributed. One of the main questions in epidemiology and public health is to
find the optimal allocation of preventive and corrective resources to contain an epidemic outbreak in a cost-optimal manner.
An identical question can be asked in the context of designing protection
strategies for other cyber-physical networks, motivating the main problem
covered in this chapter:
\begin{problem*}
Find the cost-optimal allocation of preventive and corrective
resources to protect a cyber-physical network against spreading processes.
\end{problem*}
In the field of systems reliability, there is a well-developed theory
of preventive and corrective maintenance for single components or
machines, but there is a lack of a theoretical framework to analyze
large-scale interdependent systems \cite{rausand2004system}. The state-of-the-art in the reliability
analysis of networked systems is mostly based on Markov models \cite{iyer1986analysis,reibman1991reliability,rausand2004system,dominguez2007integrated}. These
models usually suffer from scalability issues, since the state space
grows exponentially fast with the number of components under
consideration. Similar Markov models have also been proposed in the analysis
of disease spreading in networked populations. A rich and growing
literature is arising in this context, proposing a variety of approaches
to find efficient allocation of protection resources to contain an
epidemic outbreak. In a series of papers, Preciado et al. developed a mathematical framework, based on dynamic systems theory
and convex optimization, to find the optimal distribution of protection
resources in a complex network \cite{PJ13, PZEJP13,PDJ13,PZS14}. In particular,
they showed that it is possible to {\it find the cost-optimal distribution
of vaccines and antidotes in a (possibly weighted and directed) social
network of nonidentical nodes in polynomial time} using geometric
programming \cite{cones14}. This framework has also be extended
to find the {\it allocation of traffic-control resources} to find
the cost-optimal traffic profile in a transportation network to contain
the spread of a disease among cities \cite{PZ13}.

\section{Mathematical Framework}

We introduce notation and preliminary results needed in our derivations.
In the rest of the paper, we denote by $\mathbb{R}_{+}^{n}$ (respectively,
$\mathbb{R}_{++}^{n}$) the set of $n$-dimensional vectors with nonnegative
(respectively, positive) entries. We denote vectors using boldface letters
and matrices using capital letters. $I$ denotes the identity matrix
and $\mathbf{1}$ the vector of all ones. $\Re\left(z\right)$ denotes
the real part of $z\in\mathbb{C}$.

\subsection{\label{sub:Graph-Theory}Graph Theory}

A \emph{weighted}, \emph{directed} graph (also called digraph) is
defined as the triad $\mathcal{G}\triangleq\left(\mathcal{V},\mathcal{E},\mathcal{W}\right)$,
where (\emph{i}) $\mathcal{V}\triangleq\left\{ v_{1},\dots,v_{n}\right\} $
is a set of $n$ nodes, (\emph{ii}) $\mathcal{E}\subseteq\mathcal{V}\times\mathcal{V}$
is a set of ordered pairs of nodes called directed edges, and (\emph{iii})
the function $\mathcal{W}:\mathcal{E}\rightarrow\mathbb{R}_{++}$
associates \textit{positive} real weights to the edges in $\mathcal{E}$.
By convention, we say that $\left(v_{j},v_{i}\right)$ is an edge
from $v_{j}$ pointing towards $v_{i}$. We define the in-neighborhood
of node $v_{i}$ as $\mathcal{N}_{i}^{in}\triangleq\left\{ j:\left(v_{j},v_{i}\right)\in\mathcal{E}\right\} $,
i.e., the set of nodes with edges pointing towards $v_{i}$. We define
the weighted \emph{in-degree} (resp., out-degree) of node $v_{i}$
as $\deg_{in}\left(v_{i}\right)\triangleq\sum_{j\in\mathcal{N}_{i}^{in}}\mathcal{W}\left(\left(v_{j},v_{i}\right)\right)$
(resp., $\deg_{out}\left(v_{i}\right)\triangleq\sum_{j\in\mathcal{N}_{i}^{out}}\mathcal{W}\left(\left(v_{j},v_{i}\right)\right)$);
in other words, the weighted degrees are the sum of the edge weights
attached to a node.

The \emph{adjacency matrix} of a weighted, directed graph $\mathcal{G}$,
denoted by $A_{\mathcal{G}}=[a_{ij}]$, is a $n\times n$ matrix defined
entry-wise as $a_{ij}=\mathcal{W}((v_{j},v_{i}))$ if edge $(v_{j},v_{i})\in\mathcal{E}$,
and $a_{ij}=0$ otherwise \cite{Big93}. Given a $n\times n$
matrix $M$, we denote by $\mathbf{v}_{1}\left(M\right),\ldots,\mathbf{v}_{n}\left(M\right)$
and $\lambda_{1}\left(M\right),\ldots,\lambda_{n}\left(M\right)$
the set of eigenvectors and corresponding eigenvalues of $M$, respectively,
where we order them in decreasing order of their real parts, i.e.,
$\Re\left(\lambda_{1}\right)\geq\Re\left(\lambda_{2}\right)\geq\ldots\geq\Re\left(\lambda_{n}\right)$.
We call $\lambda_{1}\left(M\right)$ and $\mathbf{v}_{1}\left(M\right)$
the dominant eigenvalue and eigenvector of $M$. The spectral radius
of $M$, denoted by $\rho\left(M\right)$, is the maximum modulus
of an eigenvalue of $M$.

In this paper, we only consider graphs with positively weighted edges;
hence, the adjacency matrix of a graph is always nonnegative. Conversely,
given a $n\times n$ nonnegative matrix $A$, we can associate a directed
graph $\mathcal{G}_{A}$ such that $A$ is the adjacency matrix of
$\mathcal{G}_{A}$. Finally, a nonnegative matrix $A$ is \emph{irreducible}
if and only if its associated graph $\mathcal{G}_{A}$ is strongly
connected.

In our derivations, we use Perron-Frobenius lemma, from the theory
of nonnegative matrices \cite{meyer2000matrix}:
\begin{lemma}
\label{lem:Perron-Frobenius}(Perron-Frobenius) Let $M$ be a nonnegative,
irreducible matrix. Then, the following statements about its spectral
radius, $\rho\left(M\right)$, hold:

\emph{(}a\emph{)} $\rho\left(M\right)>0$ is a simple eigenvalue of
$M$,

\emph{(}b\emph{)} $M\mathbf{u}=\rho\left(M\right)\mathbf{u}$, for
some $\mathbf{u}\in\mathbb{R}_{++}^{n}$, and

\emph{(}c\emph{)} \textup{$\rho\left(M\right)=\inf\left\{ \lambda\in\mathbb{R}:M\mathbf{u}\leq\lambda\mathbf{u}\mbox{ for }\mathbf{u}\in\mathbb{R}_{++}^{n}\right\} $.}\end{lemma}

\begin{remark}
Since a matrix $M$ is \emph{irreducible} if and only if its associated
digraph $\mathcal{G}_{M}$ is strongly connected, the above lemma
also holds for the spectral radius of the adjacency matrix of any
(positively) weighted, strongly connected digraph.
\end{remark}
\begin{corollary}
\label{cor:Eig equals Rad}Let $M$ be a nonnegative, irreducible
matrix. Then, its eigenvalue with the largest real part, $\lambda_{1}\left(M\right)$,
is real, simple, and equal to the spectral radius $\rho\left(M\right)>0$.\end{corollary}

\subsection{\label{sub:Epidemic-Model}Stochastic Spreading Model in Arbitrary
Networks}

We formulate the simplest version of the problem under consideration
using a generalization of the SIS model,
popularly used to model spreading dynamics in networks, such as the
propagation of diseases in a networked population \cite{AL-JAY:76,CWWLF08,MOK09}
or malware in a compute network \cite{JOK-SRW:91,JOK-SRW:93,CW-JCK-MCE:00,MG-WG-DT:03}. This generalization
of the SIS model, called \emph{Heterogeneous Networked SIS model} (HeNeSIS), is
a continuous-time networked Markov process in which each node in the
network can be in one out of two possible states, namely, susceptible
or infected. In the context of systems reliability, each node
in the networked Markov process represents a component in a networked
infrastructure, and the \emph{susceptible}  and \emph{infected}
states correspond to \emph{operational} and \emph{faulty}
states of these components, respectively. Over time, each node $v_{i}\in\mathcal{V}$
in the networked Markov process can change its state according to
a stochastic process parameterized by (\emph{i}) the edge propagation
rate $\beta_{ij}$, and (\emph{ii}) its node recovery rate $\delta_{i}$.
In what follows, we shall describe the dynamics of the HeNeSIS model.

\begin{figure}
\centering
\includegraphics[width=0.7\textwidth]{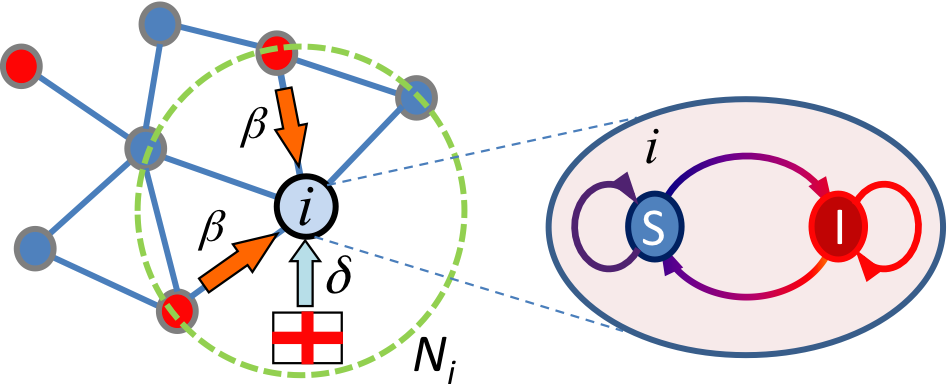}
\caption{Networked Markov process with 2 states per node, corresponding
to the HeNeSIS spreading model. Infected (resp., susceptible) nodes
are plotted in red (resp., blue).}
\end{figure}%

The dynamics of the HeNeSIS model can be described as follows. The state
of node $v_{i}$ at time $t\geq0$ is a binary random variable $X_{i}\left(t\right)\in\{0,1\}$.
The state $X_{i}\left(t\right)=0$ (resp., $X_{i}\left(t\right)=1$)
indicates that node $v_{i}$ is in the susceptible (resp., infected)
state. We define the vector of states as $X\left(t\right)=\left(X_{1}\left(t\right),\ldots,X_{n}\left(t\right)\right)^{T}$.
The state of a node can experience two possible stochastic transitions:
\begin{enumerate}
\item Assume node $v_{i}$ is in the susceptible state at time $t$. This
node can switch to the infected state during the time interval $\left[t,t+\Delta t\right)$
with a probability that depends on: (\emph{i}) the propagation rates
$\left\{ \beta_{ij},\mbox{ for }j\in\mathcal{N}_{i}^{in}\right\} $,
and (\emph{iii}) the states of its in-neighbors $\left\{ X_{j}\left(t\right),\mbox{ for }j\in\mathcal{N}_{i}^{in}\right\} $.
Formally, the probability of this transition is given by
\begin{equation}
\Pr\left(X_{i}(t+\Delta t)=1|X_{i}(t)=0,X(t)\right)=\sum_{j\in\mathcal{N}_{i}^{in}}\beta_{ij}X_{j}\left(t\right)\Delta t+o(\Delta t),\label{eq:}
\end{equation}
 where $\Delta t>0$ is considered an asymptotically small time interval.
\item Assuming node $v_{i}$ is infected, the probability of $v_{i}$ recovering
back to the susceptible state in the time interval $\left[t,t+\Delta t\right)$
is given by
\begin{equation}
\Pr(X_{i}(t+\Delta t)=0|X_{i}(t)=1,X(t))=\delta_{i}\Delta t+o(\Delta t),\label{eq:2}
\end{equation}
where $\delta_{i}>0$ is the curing rate of node $v_{i}$.
\end{enumerate}
In the context of failure propagation in networked infrastructure,
$\beta_{ij}$ represents the Poisson rate at which a failure in the
element located at node $v_{j}$ propagates to the element in node
$v_{i}$. Similarly, $\delta_{i}$ represents the Poisson rate at
which a fault at component $v_{i}$ is cleared. This
HeNeSIS model is therefore a continuous-time Markov process with $2^{n}$
states in the limit $\Delta t\to0^{+}$. Unfortunately, the exponentially
increasing state space makes this model hard to analyze for large-scale
networks. Using the Kolmogorov
forward equations and a mean-field approach \cite{MOK09}, one
can approximate the dynamics of the spreading process using a system
of $n$ ordinary differential equations, as follows. Let us define
$p_{i}\left(t\right)\triangleq\Pr\left(X_{i}\left(t\right)=1\right)=E\left(X_{i}\left(t\right)\right)$,
i.e., the probability of node $v_{i}$ being infected (or faulty)
at time $t$. Hence, the Markov differential equation \cite{van2006performance}
for the state $X_{i}\left(t\right)=1$ is the following,
\begin{equation}
\frac{dp_{i}\left(t\right)}{dt}=\left(1-p_{i}\left(t\right)\right)\sum_{j=1}^{n}\beta_{ij}p_{j}\left(t\right)-\delta_{i}p_{i}\left(t\right).\label{eq:heteroSIS}
\end{equation}
Considering $i=1,\ldots,n$, we obtain a system of nonlinear differential
equation with a complex dynamics. In the following, we derive a sufficient
condition for the spreading process to die out exponentially fast.
Let us define the vector $\mathbf{p}\left(t\right)\triangleq\left(p_{1}\left(t\right),\ldots,p_{n}\left(t\right)\right)^{T}$,
and the matrices $B_{\mathcal{G}}\triangleq\left[\beta_{ij}\right]$,
$D\triangleq\mbox{diag}\left(\delta_{i}\right)$. Notice that $B_{\mathcal{G}}$
is the weighted adjacency matrix of a weighted, directed graph with
edge-weight function $\mathcal{W}\left(v_{j}\to v_{i}\right)=\beta_{ij}$;
in other words, the weights of the directed link from $v_{j}$ to
$v_{i}$ is $\beta_{ij}$. The ODE under consideration presents an
equilibrium point at $\mathbf{p}^{*}=0$, called the disease-free (or fault-free) equilibrium.
A stability analysis of this ODE around the equilibrium provides the
following stability result \cite{PZEJP13}:

\begin{proposition} \label{prop:Heterogeneous SIS stability condition}Consider
the nonlinear HeNeSIS model in (\ref{eq:heteroSIS}) and assume $\beta_{ij},\delta_{i}>0$.
Then, if the eigenvalue with largest real part of $B_{\mathcal{G}}-D$
satisfies 
\begin{equation}
\Re\left[\lambda_{1}\left(B_{\mathcal{G}}-D\right)\right]\leq-\varepsilon,\label{eq:Spectral Control}
\end{equation}
for some $\varepsilon>0$, the disease-free equilibrium ($\mathbf{p}^{*}=\boldsymbol{0}$)
is globally exponentially stable, i.e., $\left\Vert \mathbf{p}\left(t\right)\right\Vert \leq\left\Vert \mathbf{p}\left(0\right)\right\Vert K\exp\left(-\varepsilon t\right)$,
for some $K>0$.
\end{proposition} 

\section{\label{sub:Problem-Statements}A Quasiconvex Framework for Optimal Resource Allocation}

Assume that the fault propagation and recovery rates, $\beta_{ij}$
and $\delta_{i}$, are adjustable by allocating protection resources
on the edges and nodes of the networked Markov process. We consider
two types of protection resources: (\emph{i}) preventive resources
(e.g., vaccinations in the case of disease spreading), and (\emph{ii})
corrective resources (e.g., antidotes). We assume that the propagation rate $\beta_{ij}$ can be reduced using preventive resources. Also, allocating
corrective resources at node $v_{i}$ increases the recovery rate
$\delta_{i}$. We assume that we are able to, simultaneously, modify
the fault propagation and recovery rates of $v_{i}$ within feasible
intervals $0<\underline{\beta}_{ij}\leq\beta_{ij}\leq\bar{\beta}_{ij}$
and $0<\underline{\delta}_{i}\leq\delta_{i}\leq\bar{\delta}_{i}<\Delta$,
where $\Delta$ is an uniform upper bound in the achievable recovery
rate, which is assumed to be known \emph{a priori}. The particular
values of $\beta_{ij}$ and $\delta_{i}$ depend on the amount of
preventive and corrective resources allocated at node $v_{i}$. We
consider that protection resources have an associated cost. We define
two cost functions, the prevention (or vaccination) cost function
$f_{ij}\left(\beta_{ij}\right)$ and the correction (or antidote)
cost function $g_{i}\left(\delta_{i}\right)$, that account for the
cost of tuning the fault propagation and recovery rates to $\beta_{ij}\in\left[\underline{\beta}_{ij},\bar{\beta}_{ij}\right]$
and $\delta_{i}\in\left[\underline{\delta}_{i},\bar{\delta}_{i}\right]$,
respectively.

In this context of protection design, one can study a type of resource
allocation problems, called the \emph{budget-constrained} allocation
problem. In the\emph{ budget-constrained} problem we are assigned
a total budget $C$ to invest on protection resources and we need
to find the best allocation of preventive and/or corrective resources
to maximize a measure of the network resilience. In \cite{PZEJP13} and \cite{cones14},
the authors proposed a measure of the network resilience based on
the norm of the vector of probabilities of fault probabilities, $\left\Vert \mathbf{p}\left(t\right)\right\Vert $.
In particular, the exponential rate of decay of such a vector is a
measure of the ability of the networked infrastructure to recover
from random failures. In other words, assuming that we are able to
control the system to satisfy the condition $\left\Vert \mathbf{p}\left(t\right)\right\Vert \leq\left\Vert \mathbf{p}\left(0\right)\right\Vert K\exp\left(-\varepsilon t\right)$,
the exponential decay rate $\varepsilon$ measures the ability
of the networked infrastructure to `self-heal' from random contingencies.

Based on Proposition \ref{prop:Heterogeneous SIS stability condition},
the decay rate of an epidemic outbreak is determined by $\varepsilon$
in (\ref{eq:Spectral Control}). Thus, given a budget $C$, the budget-constrained
allocation problem is formulated as follows:
\begin{problem}
\label{Problem: Budget Constrained Allocation}(\emph{Budget-constrained
allocation}) \emph{Given the following elements: (i) A directed network
$\mathcal{G}=\left(\mathcal{V},\mathcal{E}\right)$ representing failure
dependencies between components in a networked infrastructure, (ii)
a set of cost functions $f_{ij}\left(\beta_{ij}\right)$,$g_{i}\left(\delta_{i}\right)$,
}(\emph{iii})\emph{ bounds on the fault propagation and recovery rates
$0<\underline{\beta}_{ij}\leq\beta_{ij}\leq\overline{\beta}_{ij}$
and $0<\underline{\delta}_{i}\leq\delta_{i}\leq\overline{\delta}_{i}$,}
\emph{and (iv) a total budget $C$, find the cost-optimal distribution
of (preventive and corrective) protection resources to maximize the
exponential decay rate $\varepsilon$.}
\end{problem}
Based on Proposition \ref{prop:Heterogeneous SIS stability condition},
we can state this problem as the following optimization program:

\emph{
\begin{align}
\underset{{\scriptscriptstyle \varepsilon,\left\{ \beta_{ij}\right\} _{\left(j,i\right)\in\mathcal{E}},\left\{ \delta_{i}\right\} _{i=1}^{n}}}{\mbox{maximize }} & \varepsilon\label{eq:Budget-Constrained Spectral Problem}\\
\mbox{subject to } & \Re\left[\lambda_{1}\left(B_{\mathcal{G}}-D\right)\right]\leq-\varepsilon,\label{eq:Spectral constraint in budget problem}\\
 & \sum_{\left(j,i\right)\in\mathcal{E}}f_{ij}\left(\beta_{ij}\right)+\sum_{i\in\mathcal{V}}g_{i}\left(\delta_{i}\right)\leq C,\label{eq:Budget constraint in budget problem}\\
 & \underline{\beta}_{ij}\leq\beta_{ij}\leq\overline{\beta}_{ij},\mbox{ }\left(j,i\right)\in\mathcal{E};\mbox{ }\underline{\delta}_{i}\leq\delta_{i}\leq\overline{\delta}_{i},\mbox{ }i\in\mathcal{V},\label{eq:Square constraint for beta in budget problem}
\end{align}
}where (\ref{eq:Budget constraint in budget problem}) is the budget
constraint.

In the following section, we propose an approach to find the optimal
budget-constraint allocation in polynomial time for weighted and directed
contact networks, under certain convexity assumptions on the cost
functions $f_{ij}$ and $g_{i}$.

\subsection{\label{sec:Convex Framework}A Geometric Programming Approach}

We propose a convex formulation to solve the budget-constrained in
weighted, directed networks using \emph{geometric programming (GP)}
\cite{BV04}. Geometric programs are a type of quasiconvex optimization
problems that can be easily transformed into convex programs and solved
in polynomial time. We start our exposition by briefly reviewing some
concepts used in our formulation. Let $x_{1},\ldots,x_{n}>0$ denote
$n$ decision variables and define $\mathbf{x}\triangleq\left(x_{1},\ldots,x_{n}\right)\in\mathbb{R}_{++}^{n}$.
In the context of GP, a \emph{monomial $h(\mathbf{x})$} is defined
as a real-valued function of the form $h(\mathbf{x})\triangleq dx_{1}^{a_{1}}x_{2}^{a_{2}}\ldots x_{n}^{a_{n}}$
with $d>0$ and $a_{i}\in\mathbb{R}$. A \emph{posynomial} function
$q(\mathbf{x})$ is defined as a sum of monomials, i.e., $q(\mathbf{x})\triangleq\sum_{k=1}^{K}c_{k}x_{1}^{a_{1k}}x_{2}^{a_{2k}}\ldots x_{n}^{a_{nk}}$,
where $c_{k}>0$. Posynomials are closed under addition, multiplication,
and nonnegative scaling. A posynomial can be divided by a monomial,
with the result a posynomial.

A geometric program (GP) is an optimization problem of the form (see
\cite{BKVH07} for a comprehensive treatment):
\begin{align}
\mbox{minimize } & f(\mathbf{x})\label{eq:General GP}\\
\mbox{subject to } & q_{i}(\mathbf{x})\leq1,\: i=1,...,m,\nonumber \\
 & h_{i}(\mathbf{x})=1,\: i=1,...,p,\nonumber 
\end{align}
where $q_{i}$ are posynomial functions, $h_{i}$ are monomials, and
$f$ is a convex function in log-scale%
\footnote{Geometric programs in standard form are usually formulated assuming
$f\left(\mathbf{x}\right)$ is a posynomial. In our formulation, we
assume that $f\left(\mathbf{x}\right)$ is in the broader class of
convex functions in logarithmic scale.%
}. A GP is a quasiconvex optimization problem \cite{BV04} that can
be transformed to a convex problem. This conversion is based on
the logarithmic change of variables $y_{i}=\log x_{i}$, and a logarithmic
transformation of the objective and constraint functions (see \cite{BKVH07}
for details on this transformation). After this transformation, the
GP in (\ref{eq:General GP}) takes the form
\begin{align}
\mbox{minimize } & F\left(\mathbf{y}\right)\label{eq:Transformed GP}\\
\mbox{subject to } & Q_{i}\left(\mathbf{y}\right)\leq0,\: i=1,...,m,\nonumber \\
 & \mathbf{b}_{i}^{T}\mathbf{y}+\log d_{i}=0,\: i=1,...,p,\nonumber 
\end{align}
where $Q_{i}\left(\mathbf{y}\right)\triangleq\log q_{i}(\exp\mathbf{y})$ and
$F\left(\mathbf{y}\right)\triangleq\log f\left(\exp\mathbf{y}\right)$. Also, assuming that $h_{i}\left(\mathbf{x}\right)\triangleq d_{i}x_{1}^{b_{1,i}}x_{2}^{b_{2,i}}\ldots x_{n}^{b_{n,i}}$,
we obtain the equality constraint above, with $\mathbf{b}_{i}\triangleq\left(b_{1,i},\ldots,b_{n,i}\right)$,
after the logarithmic change of variables. Notice
that, since $f\left(\mathbf{x}\right)$ is convex in log-scale, $F\left(\mathbf{y}\right)$
is a convex function. Also, since $q_{i}$ is a posynomial (therefore,
convex in log-scale), $Q_{i}$ is also a convex function. In conclusion,
(\ref{eq:Transformed GP}) is a convex optimization problem in standard
form and can be efficiently solved in polynomial time \cite{BV04}.

To solve Problem \ref{Problem: Budget Constrained Allocation} using
GP, it is convenient to define the `complementary' recovery rate $\widehat{\delta}_{i}\triangleq\Delta-\delta_{i}$.
We can also define a `complementary' recovery cost function as $\widehat{g}_{i}\left(\widehat{\delta}_{i}\right)\triangleq g_{i}\left(\delta_{i}\right)=g_{i}\left(\Delta-\widehat{\delta}_{i}\right)$;
in other words, instead of defining the recovery cost in terms of
the recovery rate, $\delta_{i}$, we define it in terms of its complementary
value, $\widehat{\delta}_{i}$. Hence, Problem \ref{Problem: Budget Constrained Allocation}
can be formulated as a GP if the cost functions $f_{ij}\left(\beta_{ij}\right)$
and $\widehat{g}_{i}\left(\widehat{\delta}_{i}\right)$ are posynomials
(see \cite{BKVH07}, Section 8, for a treatment about the modeling
abilities of monomials and posynomials). Therefore, the total cost
function $\sum_{\left(j,i\right)\in\mathcal{E}}f_{ij}\left(\beta_{ij}\right)+\sum_{i\in\mathcal{V}}\widehat{g}_{i}\left(\widehat{\delta}_{i}\right)$
is also a posynomial. In \cite{cones14},
Problem \ref{Problem: Budget Constrained Allocation} is transformed into a GP, using
results from the theory of nonnegative matrices and the Perron-Frobenius
lemma. The resulting formulation is described below \cite{cones14}:
\begin{thm}
\label{thm:GP for budget constrained}Consider the following elements:
\begin{enumerate}
\item A directed graph $\mathcal{G}=\left(\mathcal{V},\mathcal{E}\right)$
representing failure dependencies in a networked infrastructure.
\item Posynomial cost functions \textup{$\left\{ f_{ij}\left(\beta_{ij}\right)\right\} _{\left(j,i\right)\in\mathcal{E}}$
and} \emph{$\left\{ \widehat{g}_{i}\left(\widehat{\delta}_{i}\right)\right\} _{i\in\mathcal{V}}$}.
\item Bounds on the failure propagation and recovery rates $0<\underline{\beta}_{ij}\leq\beta_{ij}\leq\overline{\beta}_{ij}$
and \textup{$0<\underline{\delta}_{i}\leq\delta_{i}\leq\overline{\delta}_{i}<\Delta$}.
\item A maximum budget $C$ to invest in protection resources.
\end{enumerate}
Then, the optimal allocation of protection resources on edge $\left(v_{j},v_{i}\right)$
is given by $f_{ij}\left(\beta_{ij}^{*}\right)$ and the optimal allocation
of recovery resources at node $v_{i}$ is $g_{i}\left(\Delta-\widehat{\delta}_{i}^{*}\right)$,
where \textup{\emph{$\beta_{ij}^{*}$,$\widehat{\delta}_{i}^{*}$
are the optimal solution of the following GP}}:\emph{
\begin{align}
\underset{{\scriptstyle \widehat{\lambda},\left\{ \beta_{ij}\right\} _{\left(j,i\right)\in\mathcal{E}},\left\{ u_{i},\widehat{\delta}_{i}\right\} _{i\in\mathcal{V}}}}{\mbox{minimize}} & \widehat{\lambda}\label{eq:Budget-Constrained Spectral Problem-1}\\
\mbox{subject to } & \sum_{j=1}^{n}\beta_{ij}u_{j}+\widehat{\delta}_{i}u_{i}\leq\widehat{\lambda}u_{i},\label{eq:Eigenvalu condition in spectral constraint}\\
 & \sum_{\left(j,i\right)\in\mathcal{E}}f_{ij}\left(\beta_{ij}\right)+\sum_{i\in\mathcal{V}}\widehat{g}_{i}\left(\widehat{\delta}_{i}\right)\leq C,\label{eq:budget constraint in budget constrained}\\
 & \underline{\beta}_{ij}\leq\beta_{ij}\leq\overline{\beta}_{ij},\mbox{ }\left(j,i\right)\in\mathcal{E};\mbox{ }\Delta-\overline{\delta}_{i}\leq\widehat{\delta}_{i}\leq\Delta-\underline{\delta}_{i},\mbox{ }i\in\mathcal{V},\label{eq:t trick in budget constrained}
\end{align}
}
\end{thm}
It is easy to verify that the above formulation is a GP; hence, it
can be efficiently transformed into a convex optimization program
and solved in polynomial time, \cite{cones14}. The tools presented are illustrated
with a numerical simulation involving the world-wide
air transportation network.

\subsection{Controlling Epidemic Outbreaks in a Transportation Network}

We apply the above results to the design of a cost-optimal protection
strategy against epidemic outbreaks that propagate through the air
transportation network \cite{schneider2011suppressing}. We analyze
real data from the world-wide air transportation network and find
the optimal distribution of vaccines and antidotes to prevent the
viral spreading of an epidemic outbreak. We consider the budget-constrained problems in our simulations. We
limit our analysis to an air transportation network spanning the major
airports in the world, in particular, we consider only airports having
an incoming traffic greater than 10 million passengers per year (MPPY).
There are $56$ such airports world-wide and they are connected via
$1,843$ direct flights, which we represent as directed edges in a
graph. To each directed edge $(i,j)$, we assign a `contact' weight, $a_{ji}$, equal to the number
of passengers taking that flight throughout the year\footnote{Although we could have chosen other functions of the traffic to design these contact weights, we illustrate our framework using this simple set of weights. Using a different, possibly nonlinear functions, to generate these weights do not influence the tractability of our framework.} (in MPPY units).

\begin{figure}
\centering
\includegraphics[width=0.5\columnwidth]{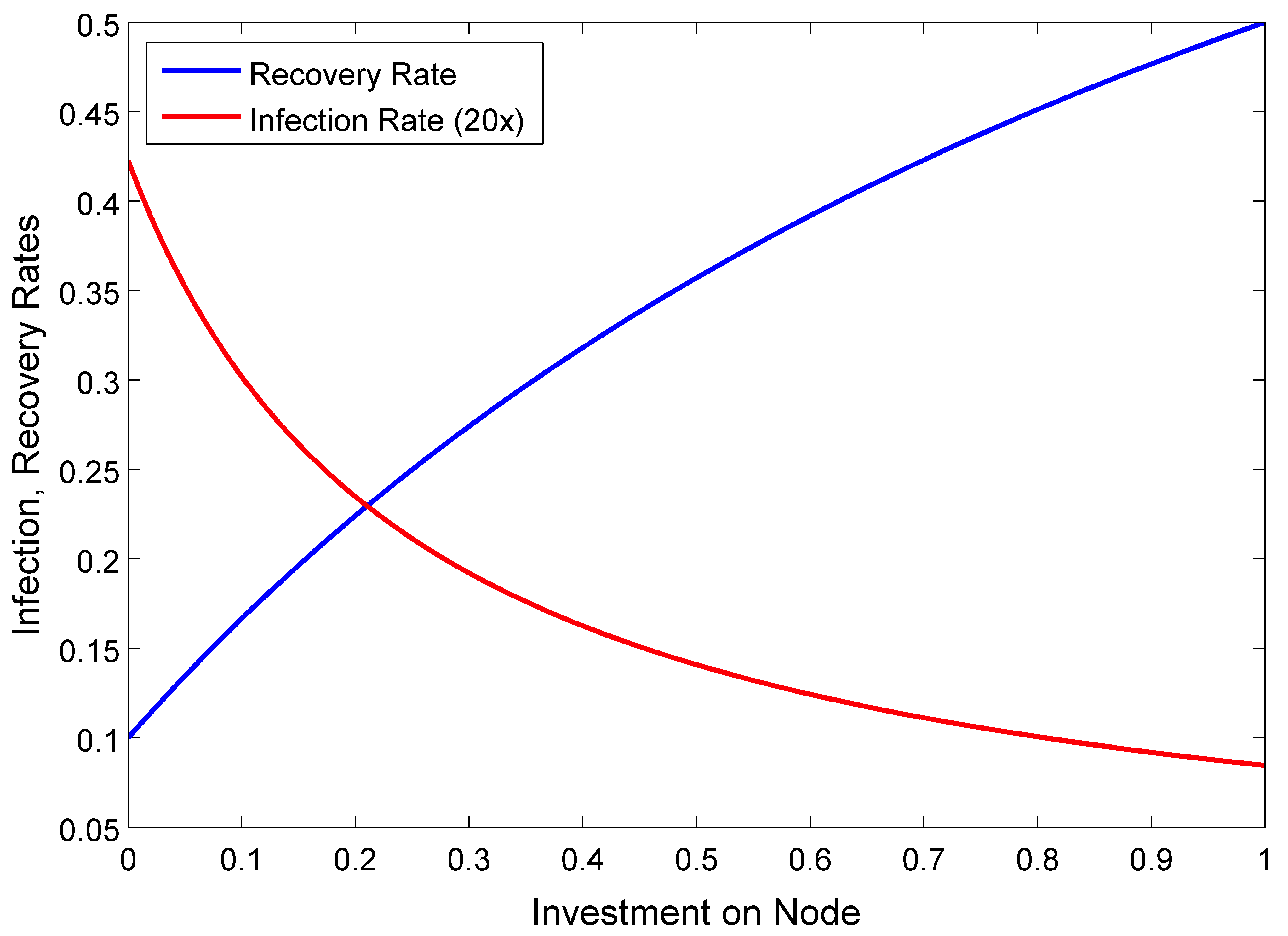}

\caption{Propagation rate (in red, and multiplied by 20,
to improve visualization) and recovery rate (in blue) achieved at
node $v_{i}$ after an investment on protection (in abscissas) is
made on that node.}\label{fig. cost functions}
\end{figure}%

In this problem, we assume that allocating preventive resources (e.g.
vaccines) at a particular airport, scale down the propagation rate
of all the incoming links in proportion to the incoming traffic. In
other words, we assume that $\beta_{ij}=\beta_{i}a_{ij}$, where $a_{ij}$
is the number of passengers per year (in MPPY) that travel from airport
$v_{j}$ to airport $v_{i}$, and $\beta_{i}$ is a scaling factor
that depends on the destination airport only. In our simulations,
we consider the following cost functions $f_{ij}\left(\beta_{ij}\right)=f\left(\beta_{i}\right)$
and $g_{i}\left(\delta_{i}\right)=g\left(\delta_{i}\right)$, where
$f$ and $g$ are the following functions (plotted in Figure \ref{fig. cost functions}):
\begin{equation}
f_{i}\left(\beta_{i}\right)=\frac{\beta_{i}^{-1}-\bar{\beta}_{i}^{-1}}{\underline{\beta}_{i}^{-1}-\bar{\beta}_{i}^{-1}},\quad g_{i}\left(\delta_{i}\right)=\frac{\left(1-\delta_{i}\right)^{-1}-\left(1-\underline{\delta}_{i}\right)^{-1}}{\left(1-\overline{\delta}_{i}\right)^{-1}-\left(1-\underline{\delta}_{i}\right)^{-1}}.\label{eq:Quasiconvex Limit-1}
\end{equation}
Notice that as we increase the amount invested on vaccines, the propagation
rate of that node is reduced from $\bar{\beta}_{i}$ to $\underline{\beta}_{i}$
(red line). Similarly, as we increase the amount invested on antidotes
at a node $v_{i}$, the recovery rate grows from $\underline{\delta}_{i}$
to $\bar{\delta}_{i}$ (blue line). Notice that both cost functions
present diminishing marginal benefit on investment.

\begin{figure}
\centering\includegraphics[width=1\textwidth]{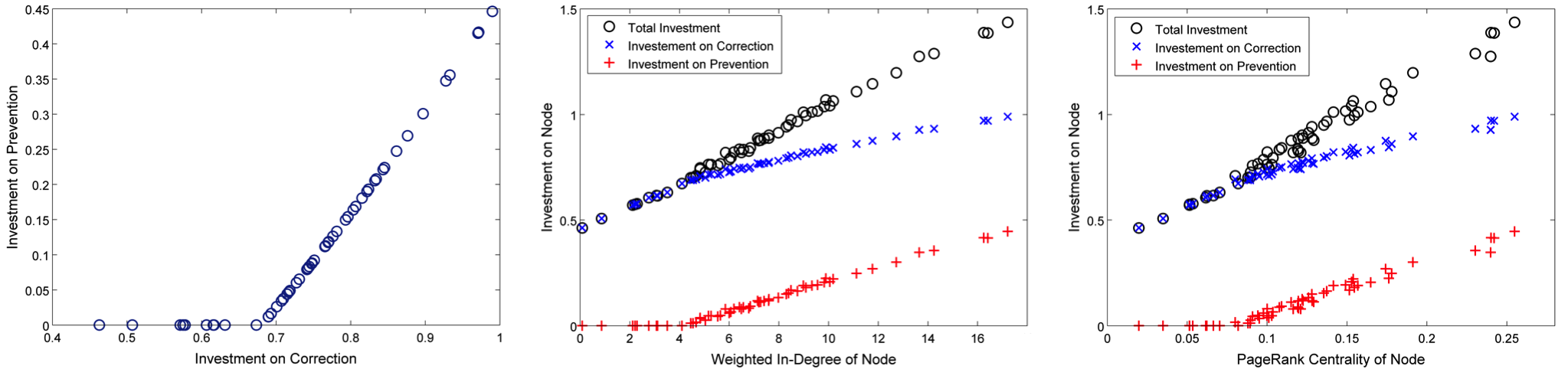}
\caption{{\small Results from the budget-constrained allocation problem. From
left to right, we have (}\emph{\small a}{\small ) a scatter plot with
the investment on correction versus prevention per node, (}\emph{\small b}{\small )
a scatter plot with the investment on protection per node and the
in-degrees, and (}\emph{\small c}{\small ) a scatter plot with the
investment on protection per node versus PageRank centralities.}\label{fig_budget-constrained} }
\end{figure}

Using the air transportation network and the cost functions specified
above, we solve the budget-constrained allocation problem using the
geometric programs in Theorems \ref{thm:GP for budget constrained}.
In the left subplot of Figure \ref{fig_budget-constrained}, we present a scatter plot
with $56$ circles (one circle per airport), where the abscissa of
each circle is equal to $g\left(\delta_{i}^{*}\right)$ and the ordinate
is $f\left(\beta_{i}^{*}\right)$, namely, the investments on allocation
of vaccines and antidotes on the airport at node $v_{i}$, for all
$v_{i}\in\mathcal{V}$. We observe an interesting pattern in the allocation
of preventive and corrective resources in the network. In particular,
we have that in the optimal allocation some airports receive only
corrective resources (indicated by circles located on top of the $x$-axis),
and some airports receive a mixture of preventive and corrective resources.
In the center and right subplots in Fig. \ref{fig_budget-constrained}, we compare the distribution
of resources with the in-degree and the PageRank%
\footnote{The PageRank vector $\mathbf{r}$, before normalization, can be computed
as $\mathbf{r}=(I-\alpha A_{\mathcal{G}}\mbox{diag}(1/\deg_{out}\left(v_{i}\right)))^{-1}\mathbf{1}$,
where $\mathbf{1}$ is the vector of all ones and $\alpha$ is typically
chosen to be $0.85$.%
} centralities of the nodes in the network \cite{New10}. In the center
subplot, we have a scatter plots where the ordinates represent investments
on prevention (red +'s), correction (blue x's), and total investment
(the sum of prevention and correction investments, in black circles)
for each airport, while the abscissas are the (weighted) in-degrees%
\footnote{It is worth remarking that the in-degree in the abscissa of Fig. \ref{fig_budget-constrained}
accounts from the incoming traffic into airport $v_{i}$ coming only
from those airports in the selective group of airports with an incoming
traffic over 10 MPPY. Therefore, the in-degree does not represent
the total incoming traffic into the airport.%
} of the airports under consideration. We again observe a nontrivial
pattern in the allocation of investments for protections. In particular,
for airports with incoming traffic less than $4$ MPPY, only corrective
resources are needed. Airports with incoming traffic over $4$ MPPY
receive both preventive and corrective resources. In the right subplot
in Fig. \ref{fig_budget-constrained}, we include a scatter plot of
the amount invested on prevention and correction for each airport
versus its PageRank centrality in the transportation network. We observe
that there is a strong correlation between the network centrality measures and the level of investment per node. In particular, there is an almost affine relationship between the total level of investment (black circles in Fig. \ref{fig_budget-constrained}, center) and the incoming traffic of an airport. Furthermore, there is a clear piece-wise linear affine relationship between the levels of investment on prevention and correction (Fig. \ref{fig_budget-constrained}, left). Similar relationships also hold when comparing the levels of investment versus the Page-Rank centralities in the airport network (Fig. \ref{fig_budget-constrained}, right).

Notice that the above distribution of protection resources correspond
to the particular cost functions chosen for our simulations. Changes
in these cost functions allow us to observe interesting phenomena
in the optimal distribution of protection resources, such as airports
with a zero protection assignment at optimality, or a distribution
of resources with a negative correlation with centrality measures.
For example, it is possible to build cases in which nodes with low
centrality (e.g. nodes with low incoming traffic and PageRank) are
assigned at optimality a higher level of protection than more central
nodes \cite{ZP14}.

\section{Towards a General Framework for Network Protection}
The framework presented in this chapter has been recently extended in several directions. In what follows, we briefly describe the following extensions: ({\emph i}) a generalized framework to cover more realistic epidemic models (beyond SIS), (\emph{ii}) a novel data-driven framework able to handle network uncertainties, and (\emph{iii}) an analysis tool that allows us to study non-Poissonian transmission and recovery rates.

\subsection{Generalized Epidemic Models}
In Nowzari et al. \cite{NPP15,NPP16a}, the authors recently studied a model of spreading, called the Generalized
Susceptible-Exposed-Infected-Vigilant (G-SEIV) model, that generalizes many
of the models in the literature, including SIS, SIR, SIRS, SEIR, SEIV, SEIS,
and SIV~\cite{BAP-DC-MF-NV-CD:10,HWH:00}. This model has two two infectious
states, called \emph{Infected} (I) and \emph{Exposed} (E), that allow us to model human behavioral changes. An individual is in the Exposed state if she is infected and contagious, but not yet aware that she is sick (i.e., in an asymptomatic incubation period). Individuals in the Infected are infected and aware of the disease, which induces a different behavior. For instance, a person knowingly infected with
a disease may have much less contact with others, yielding less chance of spreading the infection. The dynamics of this model is described below. The G-SEIV model also includes a \emph{Vigilant} (V) state, which represents healthy individuals being aware of the disease being spread. Hence, individuals in the Vigilant state are more careful in their social contacts and less likely to be infected.

\begin{figure}
\centering\includegraphics[width=0.8\textwidth]{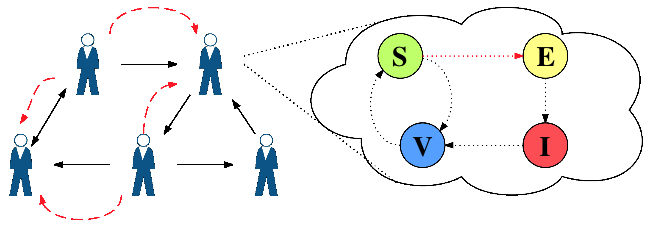}
\caption{Generalized Susceptible-Exposed-Infected-Vigilant model in a network of individuals. (Source of figure: \cite{NPP15}).}
\end{figure}

Let us denote by 
$\left[ S_i(t), E_i(t), I_i(t), V_i(t) \right]^T$ the probability vector associated with
node $i$ being in each one of these states: Susceptible, Exposed, Infected, or Vigilant, respectively. Using a mean-field approximation, the dynamics of the G-SEIV model can be described as:

\begin{align}\label{eq:continuousdynamics}
\dot{S}_i(t) &= \gamma_i V_i(t) - {\theta}_i S_i(t) - S_i(t) \hspace*{-1mm} \left( \sum_{j \in \Nin_i} {\beta}^E_i E_j(t) + {\beta}^I_i I_j(t) \hspace*{-.6mm} \right) \notag \\
\dot{E}_i(t) &= S_i(t)  \left( \sum_{j \in \Nin_i} {\beta}^E_i E_j(t) + {\beta}^I_i I_j(t)  \right) - \epsilon_i E_i(t) \notag \\
\dot{I}_i(t) &= \epsilon_i E_i(t) - {\delta}_i I_i(t) \\
\dot{V}_i(t) &= {\delta}_i I_i(t) + {\theta}_i S_i(t) - \gamma_i V_i(t) . \notag
\end{align}

Using nonlinear analysis techniques, Nowzari et al. derived the following necessary and sufficient condition for the disease to
die out exponentially fast:

\begin{theorem}\longthmtitle{Conditions for stability of disease-free equilibrium}\label{th:stability}
The disease-free equilibrium of the G-SEIV model is globally exponentially
stable if and only if the following matrix,
\begin{align}\label{eq:fullmatrix}
Q = \left[ \begin{array}{cc} TB^E A_\GG - E & TB^I A_\GG \\ E & -D \end{array} \right] .
\end{align}
is Hurwitz, where $B^E = \diag{ \beta^E }, B^I = \diag{ \beta^I }, D = \diag{ \delta }, E = \diag{ \epsilon },$  $T = \diag{  \frac{\gamma}{\theta + \gamma} }$
\end{theorem}

The above result can be used to mitigate, or
eliminate completely, the spreading of the disease. In \cite{NPP15}, the authors considered
three types of resources are available to control the disease: corrective resources (e.g., antidotes), 
preventative resources (e.g., vaccines), and preemptive resources
(e.g., awareness campaigns and/or limiting 
traffic). Under mild conditions on the cost
functions of these resources, the authors were able to bound the rate of spreading
of the undesired disease.

\subsection{Data-Driven Allocation}
\begin{figure}[t]
\centering
\includegraphics[width=0.53\textwidth]{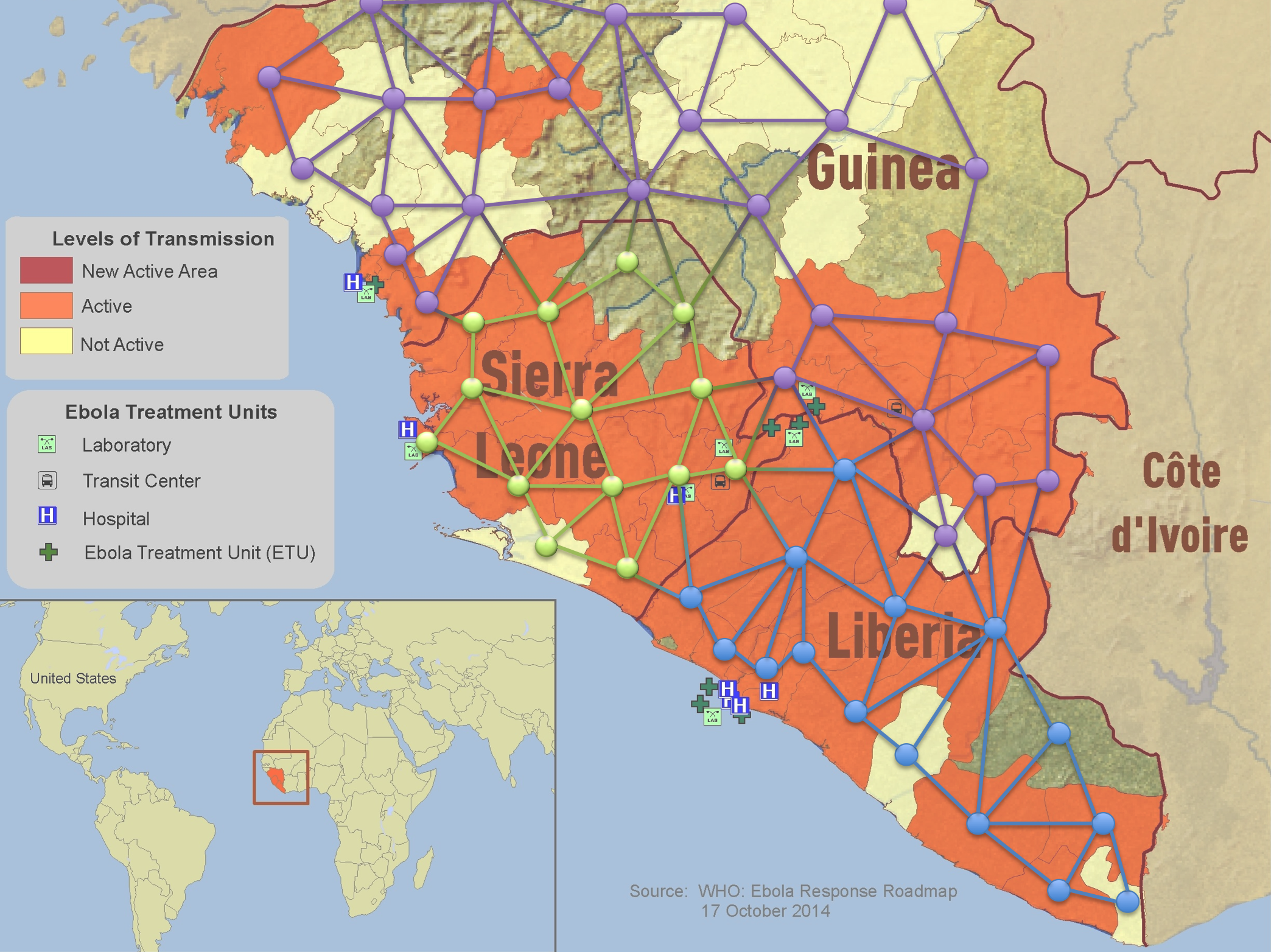}
\quad
\includegraphics[width=0.30\textwidth]{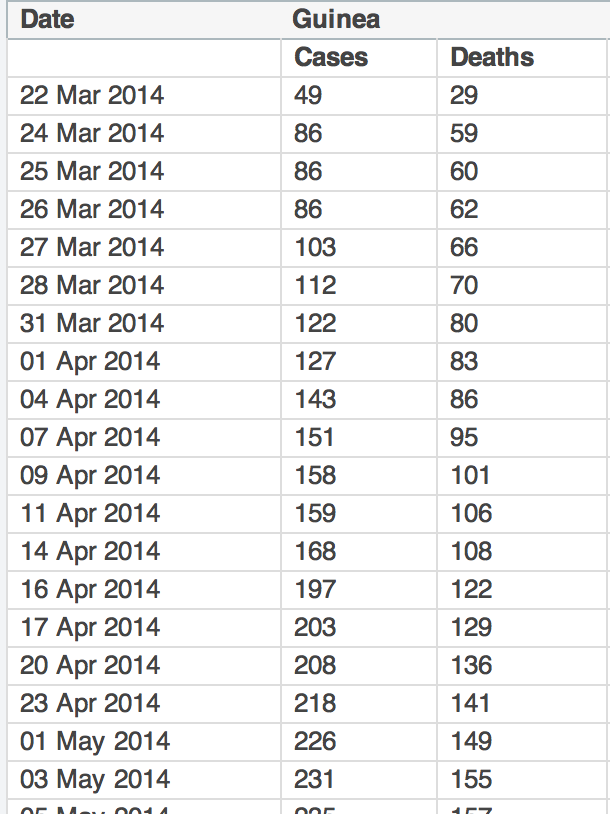}
\caption{Network of districts in West Africa affected by the 2014 Ebola outbreak. (Source: WHO)}
\label{fig_Ebola}
\end{figure}

Although current vaccination strategies assume full knowledge about the network structure and spreading rates, in most practical applications, this information is only partially known. To elaborate on this point, let us consider the following setup. Assume that each node in the network represents subpopulations (e.g., city districts) connected by edges that are determined by commuting patterns between districts. In practice, one can use traffic information and geographical proximity to infer the existence of an edge connecting districts. For example, in Fig. \ref{fig_Ebola}, we represent such a network for those districts in West Africa affected by the 2014 Ebola outbreak. On the other hand, it is very challenging to use this information to estimate the contact rates between subpopulations. Inspired by this example, we considered in \cite{HPNP15} a networked SIS model taking place in a contact network of unknown contact rates. To extract information about these unknown rates, we assumed that we have access to time series describing the evolution of the spreading process observed from a collection of sensor nodes during a finite time interval. 

In contrast to current network identification heuristics, in which a single network is identified to explain the observed data, the authors in \cite{HPNP15} developed a robust optimization framework in which an uncertainty set containing all networks that are coherent with empirical observations is defined. This characterization of the uncertainty set of networks is tractable in the context of \emph{conic geometric programming}, recently proposed by Chandrasekaran and Shah \cite{chandrasekaran2014conic}. In this context, the authors were able to efficiently find the optimal allocation of resources to control the worst-case spread that can take place in the uncertainty set of networks. In order to extract information about the contact rates, the authors considered two different sources of information that are
usually available in epidemiological problems. These sources can be classified as
(\emph{i}) \emph{prior information} about the network topology
and parameters of the disease, and (\emph{ii}) \emph{empirical observations}
about the spreading dynamics. In particular, one can consider the following
pieces of prior information: 

\begin{enumerate}

\item Assume that the sparsity pattern of the contact matrix $B_{\mathcal{G}}$
is given, although its entries are unknown. This piece of information
may be inferred from geographical proximity, commuting patterns, or
the presence of transportation links connecting subpopulations.

\item Assume that upper and lower bounds on the spreading
rates associated to each edge, i.e., $\beta_{ij}\in\left[\underline{\beta}_{ij},\overline{\beta}_{ij}\right]$,
for all $\left(i,j\right)\in\mathcal{E}$, are available. This could be inferred from
traffic densities and subpopulation sizes.

\item In practice, each district contains a large number of individuals.
Therefore, one can use the average recovery rate in the absence of
vaccination as an estimation of the nodal recovery rate. We denote
this `natural' recovery rate by $\delta_{i}^{0}$, and assume it to
be known.

\end{enumerate}

Apart from these pieces of prior information, the authors in \cite{HPNP15} also assumed that
they had access to partial observations about the evolution of the
spread over a finite time interval. In particular, assume that
we observe the dynamics of the disease for $t\in\left[0,T\right]$
from a collection of sensor nodes $\mathcal{V}_{S}\subseteq\mathcal{V}$. Based on these pieces of information, one can define an uncertainty set that contains all contact
matrices $B_{\mathcal{G}}$ consistent with both empirical observations
and prior knowledge. This set contains those contact matrices $B_{\mathcal{G}}$ such that
the transmission rates $\{\beta_{ij}\}$ are consistent with the disease dynamics.

In order to eradicate the disease at the fastest rate possible, the authors in \cite{HPNP15} considered the following control problem:
\begin{problem}
\emph{\label{prob:Main Problem}(Data-driven optimal} \emph{allocation)
Assume the following pieces of information about a viral
spread are given:}

\emph{(i) prior information about the state matrix (as described in
P1--P3);}

(\emph{ii})\emph{ a finite (and possibly sparse) data series representing
partial evolution of the spread over a set of sensor nodes $\mathcal{V}_{S}\subseteq\mathcal{V}$
during the time interval $t\in\left[T\right]$ (i.e., $\mathcal{D}$
in \eqref{eq:Data Series});}

\emph{(iii) a set of vaccine cost functions $g_{i}$ for all $i\in\mathcal{V}_{C}$,
and a range of feasible recovery rates} $\left[\underline{\delta}_{i}^{c},\overline{\delta}_{i}^{c}\right]$
such that\emph{ $1-\delta_{i}^{0}=\overline{\delta}_{i}^{c}\geq\delta_{i}^{c}\geq\underline{\delta}_{i}^{c}>0$;}

\emph{(iv) a fixed budget $C>0$ to be allocated throughout a set
of control nodes in $\mathcal{V}_{C}\subseteq\mathcal{V}$, so that
$\sum_{i\in\mathcal{V}_{C}}g_{i}(\delta_{i}^{c})\leq C$.}

\emph{Find the cost-constrained allocation of control resources to
eradicate the disease at the fastest possible exponential rate, measured
as $\rho(M(B_{\mathcal{G}},\mathbf{d}^{c}))$, over} \emph{the uncertainty
set $\Delta_{B_{\mathcal{G}}}$ of contact matrices coherent with
prior knowledge and the observations in $\mathcal{D}$.}
\end{problem}

From the perspective of optimization, Problem \ref{prob:Main Problem}
is equivalent to finding the optimal allocation of resources to minimize
the worst-case (i.e., maximum possible) decay rate $\rho(M(B_{\mathcal{G}},\mathbf{d}^{c}))$
for all $B_{\mathcal{G}}\in\Delta_{B_{\mathcal{G}}}$. In Han et al. \cite{HPNP15}, a robust optimization framework was developed to solve this problem, even in the presence of sparse observations.

\subsection{Non-Poissonian Rates}

The vast majority of spreading models over networks assume
exponentially distributed transmission and recovery rates. In
contrast, empirical observations indicate that most real-world
spreading processes do not satisfy this
assumption~\cite{Limpert2001,Lloyd2001a,Lloyd2001}. For example, the
transmission rates of human immunodeficiency viruses present a
distribution far from exponential \cite{Blythe1988}. In the context of
online social networks, empirical studies show that the rate of
spreading of information follow (approximately) a log-normal
distribution \cite{Lerman2010, Mieghem2011a}.

There are only a few results available for analyzing spreading
processes over networks with non-exponential transmission and recovery rates.
The experimental study in \cite{VanMieghem2013} confirmed the dramatic effect that non-exponential rates can have on the speed of spreading, as well as on the epidemic threshold. In~\cite{Jo2014}, an analytically solvable (although rather simplistic) model of spreading with non-exponential rates was proposed. An approximate criterion for epidemic eradication over graphs with general transmission and recovery times based on asymptotic approximations was proposed in \cite{Cator2013a}.

In the recent work \cite{OP15Phase,OP16a}, the authors  propose an alternative approach to analyze general
transmission and recovery rates using phase-type distributions. In particular, they derive
conditions for disease eradication using transmission and recovery
times that follow phase-type distributions (see, e.g.,
\cite{Asmussen1996}). The class of phase-type distributions is dense
in the space of positive-valued distributions~\cite{Cox1955}, hence,
it can be used to theoretically analyze arbitrary transmission and recovery
rates. Furthermore, there are efficient algorithms to compute the
parameters of a phase-type distributions to approximate any given
distribution \cite{Asmussen1996}. The key tool in
this analysis is a vectorial representations proposed
in~\cite{Brockett2008a}, which can be used to represent phase-type
distributions.

\section{Comparisons with Common Heuristics}
Usual approaches to distribute protection resources in a network of
agents susceptible to cascade failures are heuristics based on network centrality measures \cite{New10}.  As in the optimal framework presented in the previous section, much of the literature uses a bio-inspired epidemic models when studying harmful process with the ability to spread between interconnected agents. The main idea behind heuristic protection strategies is to rank agents according
to different measures of importance based on their location in the
network and greedily distribute protection resources based on each agents
rank. For example, Cohen et al. \cite{cohen2003efficient} proposed
a simple protection strategy called \emph{acquaintance immunization
policy} in which the most connected node of a randomly selected node
is given protective resources. This strategy was proved to be much more efficient
than random allocation of protective resources. Hayashi et al. \cite{HMM03} proposed
a simple heuristic called \emph{targeted immunization} consisting
on greedily choosing nodes with the highest degree (number of connections)
in scale-free graphs. Chung et at. \cite{chung2009distributing} studied
a greedy heuristic protection strategy based on the PageRank vector
of the contact graph. Tong et al. \cite{TPTEFC10} and Giakkoupis
et al. \cite{GGTT05} proposed greedy heuristics based on protecting
those agents that induce the highest drop in the dominant eigenvalue
of the contact graph. Recently, Prakash et al. \cite{AAITF13} proposed
several greedy heuristics to contain harmful cascades in directed
networks when nodes can be partially protected (instead of completely
removed, as assumed in previous work). These heuristics, as those
in \cite{TPTEFC10,HMM03}, are based on eigenvalue perturbation analysis.

{\it The heuristic methods in the literature are designed for a single resource type, predominantly the \textit{protective resources.} A simplified variant of the budget-constrained allocation problem is presented with only protection-type resources in order to compare the optimal solution with heuristic solutions.  }

\begin{problem} The \textbf{Network Protection Problem} is given by \label{prob}
\begin{eqnarray*}
\max_{\beta, \epsilon} && \epsilon\\
s.t. && \Re[\lambda_1(BA-\delta I)] \le -\epsilon\\
&& \sum_{i=1}^n f(\beta_i) \le C\\
&& \underline{\beta} \le \beta_i \le \bar \beta \qquad \forall i\in V.
\end{eqnarray*}
\end{problem}

\subsection{Greedy, Centrality Based Strategies}

\begin{definition}
Extract the \textbf{effective objective} in Problem \ref{prob} which is induced by the epigraph form. Define 
\begin{equation}
\epsilon({\beta}) = -\Re[\lambda_1(BA-\delta I)] \label{ee}
\end{equation}
where $B= \hbox{diag} (\beta)$ for any feasible resource allocation $\beta$.
\end{definition}
Monotonicity and continuity of the function $\epsilon({\beta})$ guarantee that fixing any feasible $\beta$ and maximizing over $\epsilon$ always causes the constraint $ \Re[\lambda_1(BA-\delta I)] \le -\epsilon$ to become tight. At the optimal point $(\beta^*,\epsilon^*)$ of Problem \ref{prob} satisfies  \[ \epsilon^*=- \Re[\lambda_1(\hbox{diag}(\beta^*)A-\delta I)]. \]  When solving the resource allocation $\beta$, $\epsilon(\beta)$ is treated as the effective objective in Problem \ref{prob}.

\begin{definition}
Define the \textbf{efficiency} of a feasible resource allocation $\beta$ as
\begin{equation}
Q(\beta) = \frac{\epsilon(\beta)- \epsilon(\bar\beta)}{\epsilon(\beta^*)-\epsilon(\bar\beta)} \in [0,1]
\end{equation}
where $\beta^*$ is a resource allocation achieving the maximum in Problem \ref{prob}.
\end{definition}
The effective objective $\epsilon(\beta)$ and the costs functions $f(\beta_i)$ are monotonically non-increasing in the resource allocations $\beta_i$ at each node, therefore $\bar\beta$ trivially achieves the minimum over the set of feasible resource allocations $\beta$.

\begin{definition}
Let $v$ be a centrality vector.  Given a budget sufficient to completely protect $k$ nodes: $C= kf(\underline\beta)$, the greedy protection strategy $\hat \beta_v$  is to completely protect $k$ nodes with the highest values in $v$. Define the protection fraction: $r=k/N$ where $N=n+m$ is the the total number of nodes.
\end{definition} 

Common centrality measures used for heuristics are degree and eigenvector centrality, \cite{HMM03}. Page rank centrality is used as in place of eigenvector centrality in the case of general digraphs, \cite{chung2009distributing}.  While Page rank depends on a parameter $\alpha$, we drop the $\alpha$ from our notation because our results hold for the whole family of Page rank vectors generated by non-trivial choices of $\alpha\in (0,1)$. \\

\begin{theorem}\label{counterex}
Given the Network Protection Problem defined in Problem \ref{prob}, with budget $C$, there exists a network $G$ satisfying 
\[Q(\hat\beta_{DEG})=Q(\hat\beta_{PR}) = 0\]
where  $r\in(0,1)$ is the fraction of nodes that can be fully protected.
\end{theorem}

\begin{wrapfigure}{o}{0.4\columnwidth}
\includegraphics[width=.4\columnwidth]{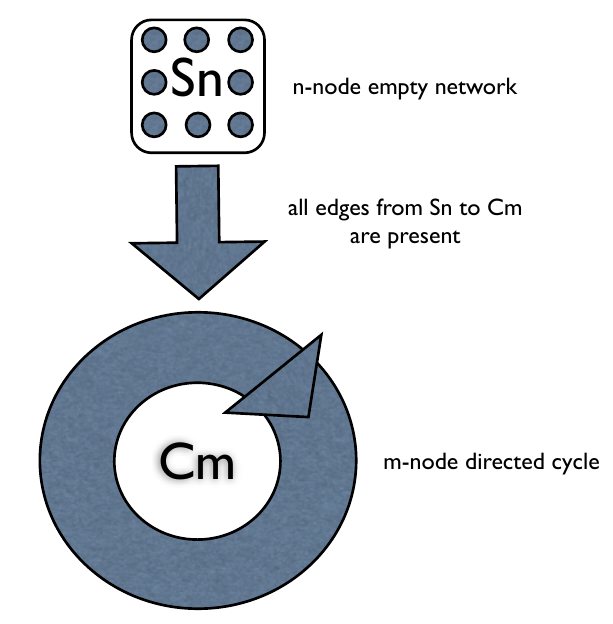}
\centering
\caption{\label{wheel}  \small{We construct the network $G$ to prove theorem \ref{counterex}.}}
\end{wrapfigure}%

Theorem \ref{counterex} is based on a worst case graph construction as shown in Fig. \ref{wheel}. Define the subgraph $C_m$ as an $m$ node directed cycle, $S_n$ as an $n$ node empty network and the there are edges from all nodes $i\in S_n$ to all nodes $j\in C_m$.  Formally, the edge set is given by
\begin{equation}
(i,j) \in \mathcal{E} \hbox{ if any of }\left\{ \begin{array}{l} i\in S_n, j\in C_m\\ i, j=i+1\in C_m\\ i=m+n, j=n+1\in C_m \end{array}\right.
\end{equation}
and in all other cases $(i,j)\not \in \mathcal{E}$. All weights are given by $\mathcal{W}(i,j)=1$ for $(i,j)\in \mathcal{E}$. Given a vaccination fraction $r$, the size of the subgraphs $C_m$ and $S_n$ must satisfy $m>n+2$ and $rN < n$ in order to generate a network for which greed heuristics have zero efficiency. Such $m$ and $n$ exist for any $r\in(0,1)$ but the networks required become very large as $r\rightarrow 1$. 

The weakly connected network $G$ results in a spreading process dominated by the nodes in $C_m$ even though nodes in $S_n$ have larger centralities. A generalization of Theorem \ref{counterex}, which builds a strongly connected network with arbitrarily small efficiency can be found in \cite{ZP14}.
 
\begin{remark}
The proof of  Theorem \ref{counterex} makes use of a constructive example for the centrality measures which identify nodes which are the most likely to fail: (a) out degree and (b) Page rank with a random walk defined as moving up the edges.  If one uses centrality measures which identify nodes which would be the most potent seeds such as (c) in degree or (d) Page rank computed using a random walk that flows down the edges, one can construct an alternative $G$ by simply reversing the direction of the edges from $S_n$ to $C_m$. Using this alternative network, one can reproduce Theorem \ref{counterex} for (c) and (d).
\end{remark}

\subsection{Greedy Heuristics and Workstation Protection}

\begin{wrapfigure}{o}{0.5\columnwidth}
\includegraphics[width=.5\columnwidth]{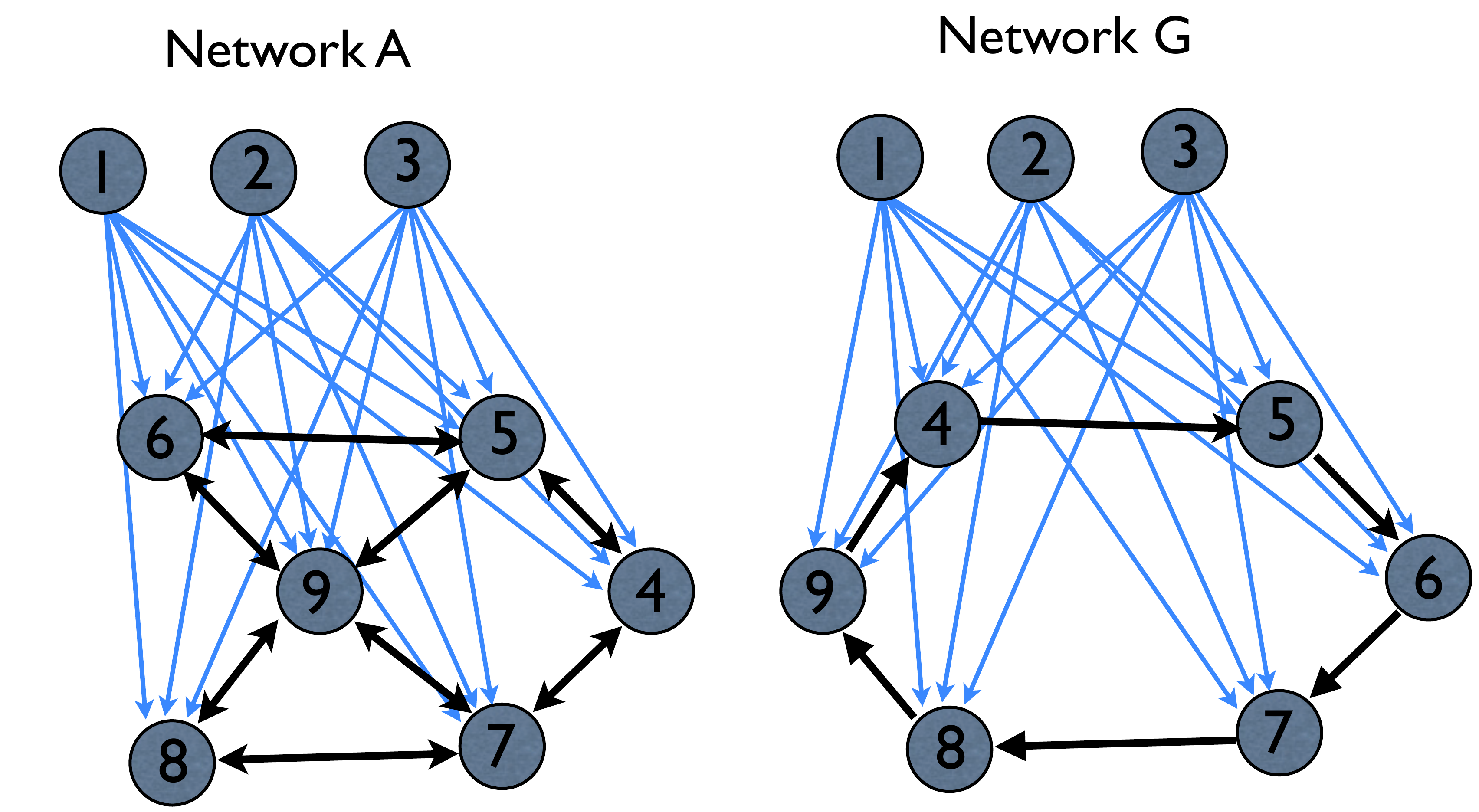}
\centering
\caption{\label{nets} \small{Network G with vertices $S_3=\{1,2,3\}$ and $C_6=\{4,5,\dots,9\}$ satisfies the conditions for the counter example network defined in Theorem \ref{counterex}.  In Network A the subgraph on $C_6$ is relaxed to be less structured for demonstration purposes.}}
\end{wrapfigure}

Consider a simple application in which such a worst case network might arise naturally: nodes are computers belonging to individuals in a work environment. Edges indicate access to files on another persons computer. \\
{\it Each workstation in $C_m$ is an element in the cyber layer, paired with one or more plants in the physical layer. Workstations in $S_n$ exist only in the cyber layer and belong to a group of administrators who can access files on all workstations in $C_m$.} \\
Workers have limited access to each others files, but do not have access to files on the administrator's computers. A virus may spread when an uninfected computer accesses an infected computer.  It is assumed that an infected workstation cannot adequately control its associated plant which leads to physical layer failures. Protection resources take the form of antivirus software with updates on a variable time interval,  software updated more frequently providing a smaller infection rate $\beta$ but updates incurring a greater cost $f(\beta)$. The cost function 

\begin{equation}
f(\beta_i) = \frac{\underline\beta(\frac{\bar \beta}{\beta_i}-1)}{\bar\beta-\underline \beta }
\end{equation}

 is chosen to satisfy $f(\bar\beta)=0$, $f(\underline\beta)=1$ and $f(\beta) \propto 1/\beta$.  This allows us to choose capacity $C$ equal to the number of nodes we wish to be able to allocate maximum protection. In our example the infection rate with outdated anti-virus software is $\bar\beta = .5$ while the maximum update rate achieves an infection rate of $\underline\beta = .01$.  Choosing a budget of $C=3$ for a network with $n=3$ and $m=6$ (such as in G or A shown in Fig. \ref{nets}), the fraction of nodes that can be maximally protected is $r=1/3$.  An infected machine has recovery rate $\delta= 0.3$, based on curative resources in the form of IT staff, which are uniformly available.

In the example, four heuristic algorithms based on greedily allocating resources with respect to centrality measures are considered.  The centrality measures are out degree, total degree, Page rank with $\alpha=.1$ and symmetrized Page rank with $\alpha=.1$. Symmetrized Page rank is computed by allowing the random walk move over a directed edge in either direction. The worst case networks are products of extreme asymmetry between $C_m$ and $S_n$, the symmetric centrality measure show that even symmetric centrality measure don't overcome the potential for arbitrarily poor behavior. 
\begin{figure}
\includegraphics[width=\textwidth]{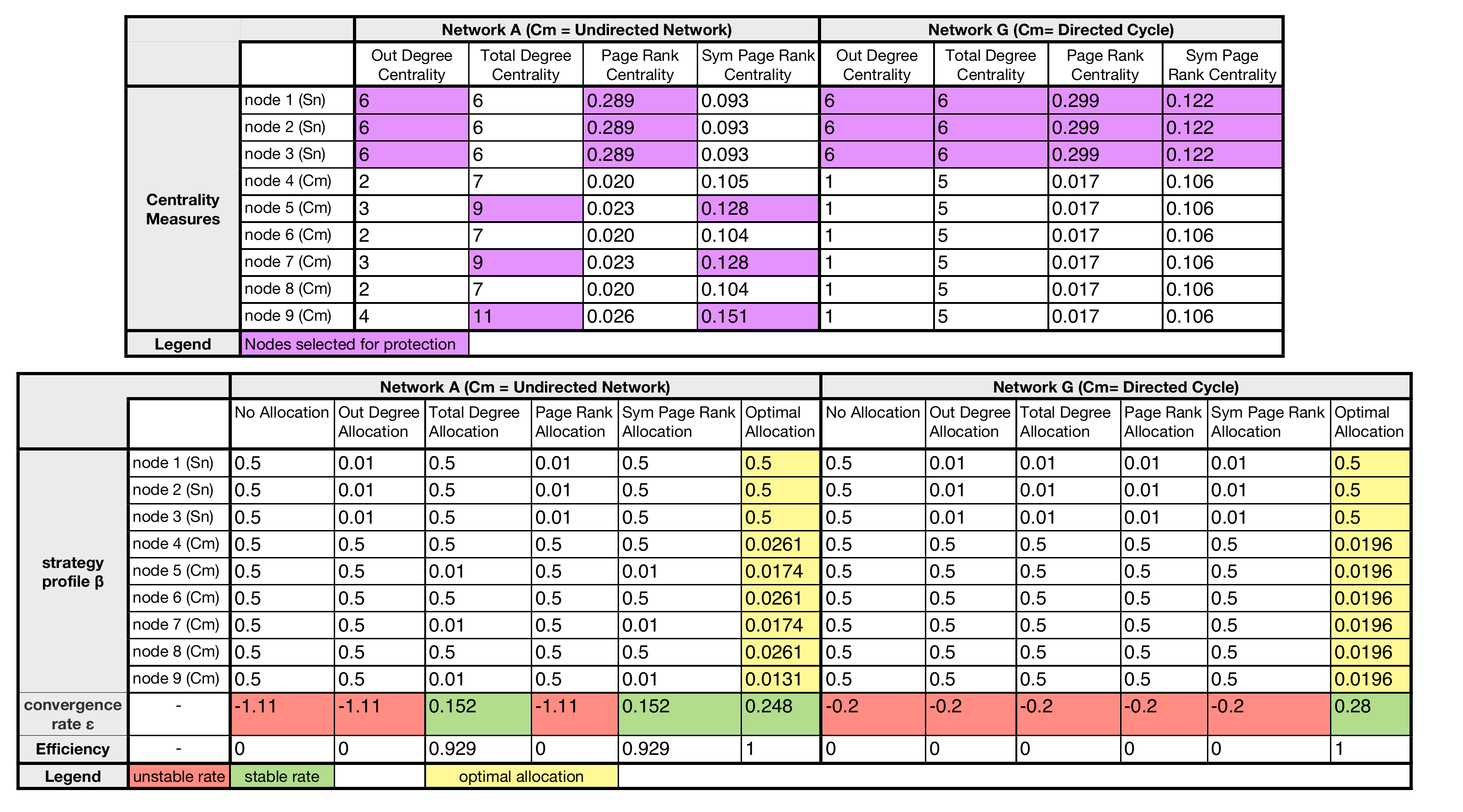}
\centering
\caption{\label{data} \small{(Top) A variety of centrality measures are used as the basis for greedy algorithms, these measures are reported for the Networks A and G. (Bottom) The allocation strategies tested are detailed, their exponential convergence rate bounds $\epsilon$ and their efficiencies are reported for comparison purposes.  For the case of the counter example network G, none of the greedy type algorithms yield a stable convergence rate.}}
\end{figure}
In Fig. \ref{data} the top table shows all of the centrality vectors for the example problem in the networks $A$ and $G$. The network $G$ is the network constructed in our analytical proofs. The network $A$ is an example of a less structured employee collaboration network which we include to demonstrate two points: (i) our constructed network G is not unique and (ii) symmetrizing heuristics are less fragile than heuristics that respect edge direction. 

In $G$ and $A$ the out degree and Page rank heuristics allocate all resources to the admins, $S_n$. This is ineffective because even though the admins are the most likely to become infected the worker group, $C_m$ cannot access their files and become infected. Furthermore, the failure of admin workstation does not lead directly to physical layer failures. Fig. \ref{data} (bottom) shows the infection rate profiles generated by the various heuristics and the optimal solution. A strategy is ineffective if the convergence rate epsilon is negative because this corresponds to unstable dynamics, where the computer virus is spreading faster than the IT staff can repair workstations.  The result is a complete failure to consistently control any of the plants in the physical layer of the system. 

\section{Conclusions}

We have studied the problem of allocating protection resources in
weighted, directed networks to contain spreading processes, such as
the propagation of viruses in computer networks, cascading failures
in complex technological networks, or the spreading of an epidemic
in a human population. We have considered two types of protection
resources: (\emph{i}) \emph{Preventive} resources able to `immunize'
nodes against the spreading (e.g. vaccines), and (\emph{ii}) \emph{corrective}
resources able to neutralize the spreading after it has reached a
node (e.g. antidotes). We assume that protection resources have an
associated cost and have then studied the \emph{budget-constrained allocation problem}, in which we find
the optimal allocation of resources to contain the spreading given
a fixed budget. We have solved this
optimal resource allocation problem in \emph{weighted and directed}
networks of \emph{nonidentical} agents in polynomial time using Geometric
Programming (GP). Furthermore, the framework herein proposed allows
\emph{simultaneous }optimization over both preventive and corrective
resources, even in the case of cost functions being node-dependent.

We have illustrated our approach by designing an optimal protection
strategy for a real air transportation network. We have limited our
study to the network of the world's busiest airports by passenger
traffic. For this transportation network, we have computed the optimal
distribution of protecting resources to contain the spread of a hypothetical
world-wide pandemic. Our simulations show that the optimal distribution
of protecting resources follows nontrivial patterns that cannot, in
general, be described using simple heuristics based on traditional
network centrality measures.

We then presented the following recent extensions on this work: ({\emph i}) a generalized framework to cover more realistic epidemic models, (\emph{ii}) a novel data-driven framework able to handle network uncertainties, and (\emph{iii}) an analysis tool that allows us to study non-Poissonian transmission and recovery rates. We concluded this chapter with a comparison between our results and common heuristics used in the literature.

\newpage

\section*{Exercises}

Consider the following three networks
with $n$ nodes:

\begin{figure}[h]
\centering
\includegraphics[width=0.9\textwidth]{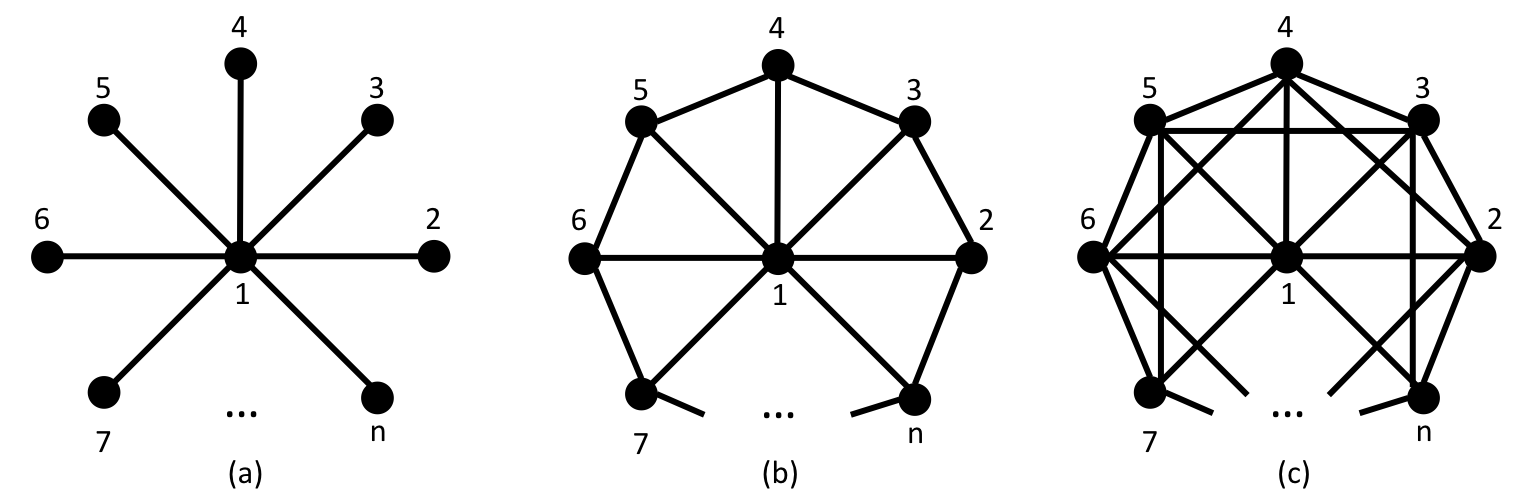}
\end{figure}

Answer the following questions:

\bigskip

{\it Question 1}. Compute the largest eigenvalue of the adjacency matrices of the
graphs in figures (a), (b) and (c) as a function of $n$.

\bigskip

{\it Question 2}. Consider the SIS model of spreading with $\beta=0.1$ and $n=100$.
For what values of $\delta$ does an epidemics die out in for each one of the three networks above?
(Reminder: The epidemic dies out when $\lambda_{1}<\delta/\beta$).

\bigskip

{\it Question 3}. Imagine you work for a health agency responsible for controlling an epidemic taking place in the above networks. Assume you can tune the spreading rates of the edges within a feasible interval $[\underline{\beta},\overline{\beta}]$. Assume the cost associated with tuning $\beta$ is given by $f_{ij}(\beta)=1/\beta$. Write the associated optimization problem for each one of the above networks. (Hint: Your answer should look like equations \ref{eq:Budget-Constrained Spectral Problem}--\ref{eq:Square constraint for beta in budget problem}).

\bigskip

{\it Question 4}. Transform the optimization problem in Question 3 into a standard geometric program. (Hint: Your answer should look like equations \ref{eq:Budget-Constrained Spectral Problem-1}--\ref{eq:t trick in budget constrained}).

\bigskip

{\it Question 5}. Implement the geometric program from Question 4 using MATLAB's CVX Toolbox \cite{cvx}.

\bibliographystyle{acm}
\bibliography{epidemics,ViralSpread}

\end{document}